\documentclass[final,5p,times,twocolumn]{elsarticle}

\usepackage{array}
\usepackage{amsmath}
\usepackage{amssymb}
\usepackage{graphicx}
\usepackage{subfigure}
\usepackage{textcomp}
\usepackage{xcolor}
\usepackage{bm}
\usepackage{caption}
\usepackage[noend]{algpseudocode}
\usepackage{algorithmicx,algorithm}
\usepackage{amsthm}

\newtheorem{proposition}{Proposition}

\usepackage{stfloats}

\journal{Fundamental Research}

\begin{document}

\begin{frontmatter}



\title{Synesthesia of Machine (SoM)-Driven Analog Precoder Optimization\\for Enhanced ISAC Performance in Sub-THz Systems}

 \author[pku]{Zonghui Yang}
 \author[gangkeguang]{Shijian Gao}
 \author[pku]{Xiang Cheng}
 \affiliation[pku]{organization={Key Laboratory of Advanced Optical Communication Systems and Networks, School of Electronics},
             addressline={Peking University},
             city={Beijing},
             postcode={100871},
             country={China}}
\affiliation[gangkeguang]{organization={Internet of Things Thrust},
             addressline={The Hong Kong University of Science and Technology (Guangzhou)},
             city={Guangzhou},
             postcode={511400},
             country={China}}

\begin{abstract}

Integrated sensing and communication (ISAC) is anticipated to be widely used in future sub-terahertz (sub-THz) systems. With the line-of-sight (LoS) propagation characteristics of sub-THz channels, ISAC transmitter design largely parallels analog precoder optimization. However, balancing both sensing and communication functionalities is challenging due to the beam squint effect in sub-THz systems, limiting ISAC performance gains. To overcome this, the unique design flexibility of sub-THz analog hardware is explored to better adapt to the electromagnetic characteristics of sub-THz channels. It is demonstrated that adjusting the equivalent channel through the analog precoder enhances dual-functional gains. Based on this, a near-optimal benchmark for analog precoder optimization is proposed. To address excessive algorithmic complexity, inspiration is drawn from the synesthesia of machine (SoM) to develop a lightweight complex-valued squint-aware network (CSP-Net). This network reduces complexity by utilizing both communication and sensing channel data, with an architecture tailored to specific data and task characteristics. The effectiveness of the proposed schemes is validated through simulations.

\end{abstract}



\begin{keyword}
Integrated sensing and communication, synesthesia of machine, sub-terahertz, analog precoding, beam squint, complex-valued neural network
\end{keyword}

\end{frontmatter}



\section{Introduction}

The future mobile information network is expected to support emerging applications such as vehicular networks and the low-altitude economy, where high-speed data transmission and ubiquitous, accurate environmental sensing are core enabling technologies \cite{survey_JCR}. Integrated sensing and communication (ISAC) not only reduces the costs by sharing hardware and signals, but also facilitates mutual assistance between communication and sensing, positioning it as a key development direction for future information networks \cite{cheng2022integrated, overview1}.
To achieve higher data rates, larger antenna arrays operating at higher frequencies have been adopted \cite{GBM}, leading to the development of sub-terahertz (sub-THz) communication systems. Meanwhile, these sub-THz systems enhance the resolution of distance and angle, making ISAC in sub-THz systems a promising avenue \cite{overview_THz_ISAC}.

Due to the severe path loss in signal propagation, the sub-THz band channel is primarily dominated by line-of-sight (LoS) paths \cite{THz_channel_com}.
Thus, the sub-THz system design can be approximated as the analog precoding design, where the beams are flexibly adjusted by the phase shifters (PSs) to enhance the array gain, thereby ensuring ISAC performance, as operated in \cite{multibeam, GC, iotj}.
However, in these sub-THz systems, the beam squint phenomenon becomes pronounced \cite{THz_bs, how_many_squint}, significantly reducing the antenna array gain and deteriorating the communication performance in existing precoding designs. To mitigate this issue, the true time delay (TTD) has been introduced in the analog hardware, providing frequency-dependent phase shifts that ensure beam alignment across subcarriers \cite{DPP}. Conversely, for sensing tasks, beam squint can lead to wider coverage and enhanced frequency-aware angle estimation capabilities. For instance, studies in \cite{squint_sensing, CBS} worked on controlling the pattern of beam squint through TTDs, facilitating user localization or target angle estimation.
Nevertheless, current sub-THz systems lack a joint design for both functionalities, failing to achieve optimal trade-offs in the presence of beam squint.

It is worth noting that the correlation between communication and sensing channels affects the ISAC performance when both functionalities co-exist \cite{IT_ISAC_Liufan}. Previous studies such as \cite{ISAC_channel_model_SD} evaluated this correlation from the perspective of power sharing between the propagation paths in sensing and communication channels. However, they do not provide direct guidance for analog precoding design. Moreover, the existing analog precoders fail to actively manipulate the degree of correlation between channels. \cite{RIS_rotation_ISAC} improved the channel correlation through external reconfigurable intelligent surfaces, thereby enhancing ISAC performance. However, these methods are primarily limited to narrowband systems and may exacerbate beam squint effects in sub-THz systems. In fact, analog precoding systems possess their own design degrees of freedom that can be leveraged to establish the performance trade-off, which has not been thoroughly explored.  Notably, the TTDs are used to adjust the beam pointing across subcarriers, holding the potential in inherently tuning the channel correlation.

Inspired by this, in this paper, we aim to integrate the unique electromagnetic propagation characteristics of sub-THz channels with the specific hardware degrees of freedom in sub-THz analog systems to optimize the analog precoder, without altering the existing sub-THz hardware architecture. First, we demonstrate the enhancement effect of the communication-sensing (C-S) channel correlation on the dual-functional gain. Based on this, we propose an optimization-based benchmark for analog precoder design to achieve the ISAC Pareto boundary. 
While this benchmark addresses the problem with near-optimal performance, the iterative and exhaustive search processes involve high complexity particularly under large arrays, probably leading to high latency in mobile networks. Inspired by research on the synesthesia of machines (SoM) \cite{overview_Som, TCOM}, which utilizes multi-modal environmental observations to enhance the transceivers design with lowered overhead, we further propose a learning-aided scheme for ISAC analog precoding design with reduced complexity. Unlike existing learning-aided precoding designs such as \cite{learning_ISAC_Elbir, learning_ISAC_precoding}, we tailor the network architecture for the analog precoding task and adopt a complex-valued network to effectively process the channel data for analog precoding design. Numerical experiments have demonstrated the superiority of the proposed schemes in terms of the dual-functional performance enhancement.

\newcommand{\RNum}[1]{\uppercase\expandafter{\romannumeral #1\relax}}

The remainder of this paper is organized as follows. Section 2 introduces the system model and the performance metrics for ISAC. Section 3 exhibits the squint-aware ISAC optimization algorithm as a near-optimal benchmark, and Section 4 presents the proposed low-complexity learning-aided approach for the analog precoder design. Section 5 introduces the benchmark scheme. Section 6 contains our experimental findings, and Section 7 concludes our work.

\textit{Notation}: $a$, $\bm a$ and $\bm A$ represent a scalar, a vector and a matrix respectively. $(\cdot)^{\text{T}}$, $(\cdot)^{\text{H}}$, $\left\|\cdot\right\|_2$, $\left\|\cdot\right\|_{\text{F}}$ denote the transpose, the conjugate transpose, 2-norm and Frobenius-norm respectively. $\odot$ and $\otimes$ denote the Hadamard product and the Kronecker product of two matrices. $\mathcal{U}(a,b]$ denotes the uniform distribution between $a$ and $b$. $\mathcal{CN}(m,\sigma^2)$ represents the complex Gaussian distribution whose mean is $m$ and covariance is $\sigma^2$. $\mathbb{E}(\cdot)$ denotes the expectation. $\bigcup$ denotes the union of sets and $\backslash$ denotes the difference set.
$\mathbb{R}$ and $\mathbb{C}$ denote the sets of real numbers and complex numbers. $\text{Re}$ and $\angle$ denote the real part and the phase of the complex numbers. $\text{diag}(\bm a)$ and $\text{diag}(\bm A)$ denote the vector diagonal matrixization operation and the operation of extracting diagonal elements from a matrix.

\vspace{-0.12cm}
\section{System Model}
\vspace{-0.18cm}
We consider a wideband multi-antenna ISAC base station (BS) serving a single-antenna user in the downlink while sensing $K$ nearby targets. The BS is equipped with a uniform planar array (UPA) comprising $N_{t}=N_{h}\times N_{v}$ antennas, where $N_{h}$ and  $N_{v}$ denote the number of antennas in horizontal and vertical directions. The ISAC signal is transmitted via orthogonal frequency division multiplexing (OFDM) with $M$ subcarriers. The frequency at the $m$-th subcarrier is $f_{m}=f_{\text{c}}+\frac{B}{M}(m-\frac{M+1}{2})$, where $f_{\text{c}}$ and $B$ denote the central frequency and the bandwidth.

\begin{figure*}[t]
  \vspace{-0.2cm}
  \setlength{\abovecaptionskip}{-0cm} 
  \setlength{\belowcaptionskip}{0.1cm} 
  \centering
 \includegraphics[width=0.92\linewidth]{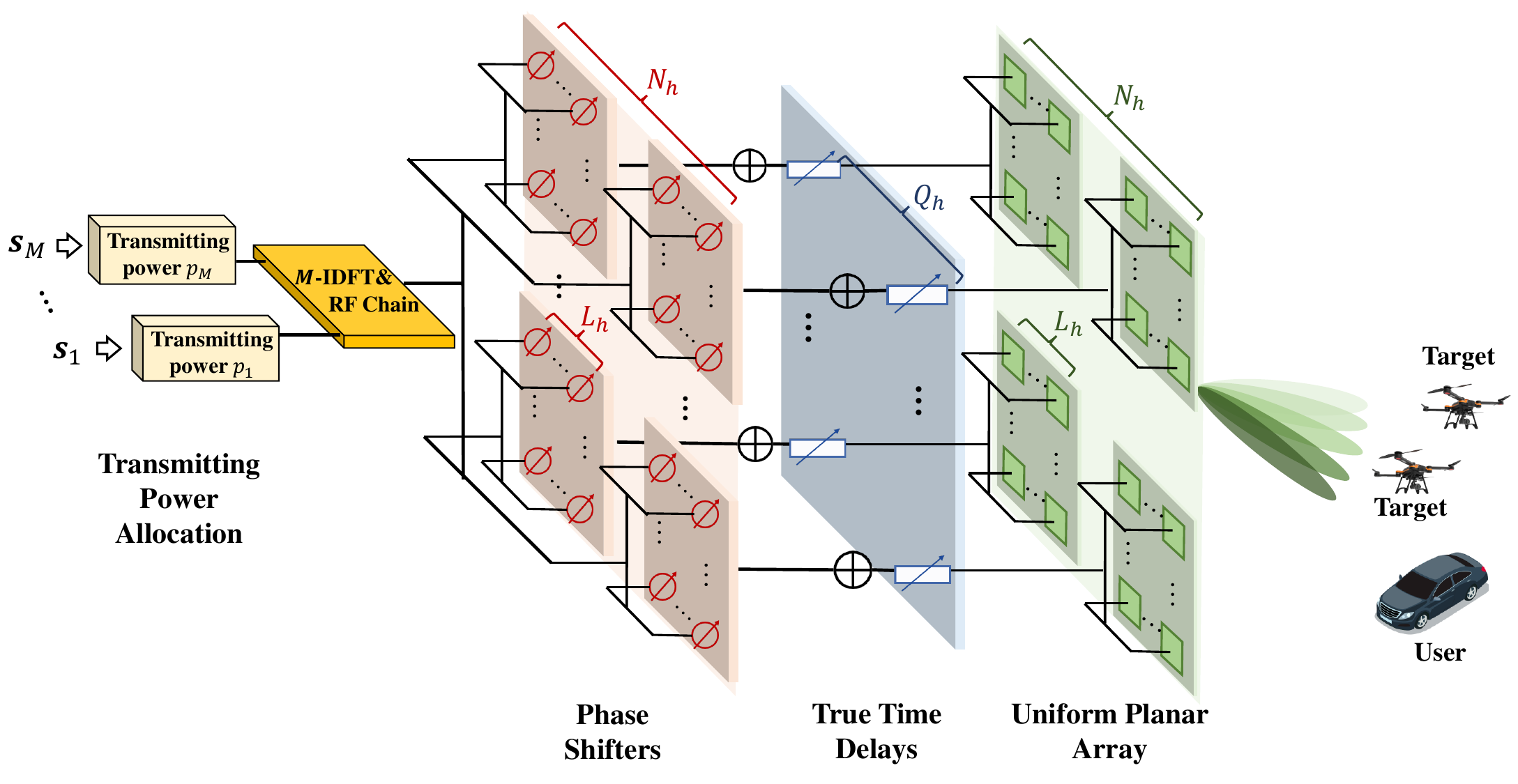}
  \vspace{0.1cm}
  \captionsetup{font=small}
  \caption{A diagram of the analog precoder structure in sub-THz systems. The baseband frequency-domain signal $\bm s$ is converted into the time domain via IDFT after transmitting power allocation. The signal is then passed through the RF chain, processed sequentially by TTDs and PSs, and transmitted from the uniform planar array to communicate with the user while simultaneously sensing the targets.}
  \label{fig:precoding structure}
  \vspace{-0.4cm}
\end{figure*}

\vspace{-0.1cm}
\subsection{Transmitted Signal in Sub-THz Systems}
\vspace{-0.0cm}
As shown in Fig.~\ref{fig:precoding structure}, we utilize a typical analog precoding structure in the sub-THz system, where a small number of TTDs are positioned between the PSs and the antenna array. This setup enables active, frequency-dependent adjustment of phase shifts to effectively deal with beam squint effects. 
In this architecture, the UPA is divided into $Q_{t}=Q_{h}\times Q_{v}$ sub-arrays, each of which is connected to a TTD. Each subarray consists of $L=L_{h}\times L_{v}$ elements, with $L_{h}=\frac{N_{h}}{Q_{h}}$ and $L_{v}=\frac{N_{v}}{Q_{v}}$. The TTDs are also partially connected with the PSs in the same way. Consequently the transmitted signal at the $m$-th subcarrier is
\vspace{-0.25cm}
\begin{equation}
    \bm x_{m}=\sqrt{p_{m}}\bm F_{\text{TD},m}\bm f_{\text{PS}}s_{m},
    \vspace{-0.2cm}
\end{equation}
where $s_{m}\!\sim\!\mathcal{CN}(0, 1)$ is the data symbol for the user at the $m$-th subcarrier. $\bm p=[p_{1}, \cdots,p_{M}]$ is the transmitting power across subcarriers. $\bm f_{\text{PS}}\in \mathbb{C}^{N_{t}}$ denotes frequency-independent phase shifters network, with $\left|\bm f_{\text{PS}} [n]\right|=1$ for each phase shifter (PS).
$\bm F_{\text{TD},m}=\text{diag}\left[\text{vec}\left( e^{j2\pi f_{m}(\bm T \otimes \bm 1^{L_{h}\times L_{v}}) } \right) \right ]$ denotes the frequency-dependent phase shifts introduced by the TTDs at the $N_{t}$ antennas, with $\bm T\in \mathbb{R}^{Q_{h}\times Q_{v}}$ being the value of delays in TTDs, and $\bm 1^{L_{h}\times L_{v}}$ being the all-one matrix.
Taking consideration of the prohibitive cost of arbitrary time delays, finite-value finite-resolution $B_{t}$-bit TTDs are utilized. i.e.,
\vspace{-0.2cm}
\begin{equation}
     \bm T[q_{h},q_{v}]\in \mathcal{T}\triangleq \left\{ \frac{b}{2^{B_{t}}}t_{\text{max}} \vert b=0,1,2,\cdots,2^{B_{t}}-1  \right\},
\vspace{-0.2cm}
\end{equation}
where $q_{h}\in[1, Q_{h}]$, $q_{v}\in[1, Q_{v}]$, and $t_{\text{max}}$ is the maximal achievable time delay determined by the hardware.

\begin{table}[t]
    \centering
    \caption{Notations related to the ISAC system.}
    \vspace{-0.15cm}
    \small
    \begin{tabular}{|p{2.05cm}| p{6.00cm}|}
    \hline
       ~~~~Notation  &  ~~~~~~~~~~~~~~~~~~~~~~~Definition \\
       \hline
       \hline
       ~~$\bm T\in \mathbb{C}^{ Q_{h}\times Q_{t}}$  &  Value of the true time delays\\
       \hline
       ~$\bm F_{\text{TD,m}}\!\in \!\mathbb{C}^{N_{t}\times N_{t}}$  &  Frequency-dependent phase shift induced by $\bm T$ \\
       \hline
       ~~~$\bm \varphi\in \mathbb{R}^{ N_{t}\times 1}$  &  Value of the phase shifters \\
       \hline
       ~~$\bm f_{\text{PS}}\in \mathbb{C}^{ N_{t}\times 1}$  &  Frequency-independent phase shift induced by $\bm \varphi$ \\
       \hline
       ~~~$\bm p\in \mathbb{R}^{M\times 1}$  &  Power allocation across subcarriers  \\
       \hline
       ~~$\bm h_{\text{c},m}\in \mathbb{C}^{N_{t}\times 1}$  &  Communication channel at the $m$-th subcarrier \\
       \hline
       ~~$\bm G_{m}\in\mathbb{C}^{N_{r}\times N_{t}}$  &  Sensing channel at the $m$-th subcarrier \\
       \hline
       ~~$\widetilde{\bm h}_{\text{c},m}\in \mathbb{C}^{N_{t}\times 1}$  &  Equivalent communication channel modified by TTDs at the $m$-th subcarrier \\
       \hline
       ~~$\widetilde{\bm G}_{m}\in\mathbb{C}^{N_{r}\times N_{t}}$  &  Equivalent sensing channel modified by TTDs at the $m$-th subcarrier \\
       \hline
       ~~$\bm D_{t}\in \mathbb{C}^{N_{t}\times N_{D}}$  &  Uniformly-sampled dictionary in beamspace\\
       \hline
       ~~~$\widehat{\bm h}^{b}_{\text{c}}\in \mathbb{R}^{{N_{D}}\times 1}$ &  Beamspace distribution of communication\\
       \hline
       ~~~$\widehat{\bm h}^{b}_{\text{s}}\in \mathbb{R}^{{N_{D}}\times 1}$ &  Beamspace distribution of sensing\\
       \hline
    \end{tabular}
    \label{tab:notation}
    \vspace{-0.3cm}
\end{table}

\vspace{-0.1cm}
\subsection{Communication Channel}
\vspace{-0.1cm}
Based on the channel model in sub-THz frequency band \cite{THz_channel_com}, the communication channel between the BS and the user at the $m$-th subcarrier is expressed as
\vspace{-0.15cm}
\begin{equation}
    \bm h_{\text{c},m}=\beta_{\text{c},m} e^{-j2\pi f_{m}\tau_{\text{c}}}\bm a_{t}(\theta_{\text{c}}, \phi_{\text{c}}, f_{m}),
    \label{equ:com_channel}
    \vspace{-0.15cm}
\end{equation}
where $\beta_{\text{c},m}\sim\mathcal{CN}(0, \sigma_{\beta}^{2})$ and $\tau_{\text{c}}\sim\mathcal{U}(0, \tau_{\text{max}}]$ are the complex gain and the propagation delay of the line-of-sight (LoS) path. $\theta_{\text{c}}\in [0,\pi]$ and $\phi_{\text{c}}\in [-\pi/2,\pi/2]$ are the elevation and the azimuth angles of the user respectively. $\bm a_{t}(\theta_{\text{c}}, \phi_{\text{c}}, f_{m})=\bm a^{h}(\theta_{\text{c}}, \phi_{\text{c}}, f_{m})  \otimes \bm a^{v}(\theta_{\text{c}}, f_{m})\in \mathbb{C}^{N_{t}\times 1}$ denotes the steering vector at the subcarrier $f_{m}$ from the transmitter towards $\theta_{\text{c}}$ and $\phi_{\text{c}}$, where
\vspace{-0.2cm}
\begin{align}
    \bm a^{h}(\theta, \phi, f_{m})[n_{h}]&=\frac{1}{\sqrt{N_{h}}} e^{j\pi \frac{f_{m}}{f_{\text{c}}}\sin{\phi}\sin{\theta}(n_{h}-1)}, \\
    \bm a^{v}(\theta, f_{m})[n_{v}]&=\frac{1}{\sqrt{N_{v}}} e^{j\pi \frac{f_{m}}{f_{\text{c}}}\cos{\theta}(n_{v}-1)}.
\end{align}
Then the received signal at the user at the $m$-th subcarrier is
\vspace{-0.2cm}
\begin{equation}
y_{\text{c},m}=\sqrt{p_{m}}\bm h_{\text{c},m}^{\text{H}}\bm F_{\text{TD},m}\bm f_{\text{PS}} s_{m}+n_{\text{c},m},
\vspace{-0.2cm}
\end{equation}
where $n_{\text{c},m}$ is additive white Gaussian noise with power being $\sigma_{\text{c}}^{2}$. The data rate for the user at the $m$-th subcarrier is 
\vspace{-0.2cm}
\begin{equation}
    R_{\text{c},m}=\log_{2}\left(1+ \frac{p_{m}\vert \bm h_{\text{c},m}^{\text{H}}\bm F_{\text{TD},m}\bm f_{\text{PS}}\vert^{2}}{\sigma_{\text{c}}^{2}} \right),
    \vspace{-0.2cm}
\end{equation}
and the total rate is $R_{\text{c}}=\sum_{m=1}^{M}R_{\text{c},m}$, indicating the communication performance for this sub-THz systems.

\vspace{-0.1cm}
\subsection{Sensing Channel}
For sensing task, the target response matrix at the $m$-th subcarrier can be written as
\vspace{-0.2cm}
\begin{equation}
\begin{aligned}
    \bm G_{m}&=\sum_{k=1}^{K}\alpha_{k,m}\bm a_{r}(\theta_{\text{s},k}, \phi_{\text{s},k}, f_{m})\bm a^{\text{T}}_{t}(\theta_{\text{s},k}, \phi_{\text{s},k}, f_{m})\\
    &=\bm A_{r,m} \bm\Sigma_{m}\bm A_{t,m}^{T},
    \label{equ:sensing_channel}
\end{aligned}
\end{equation}
where $\alpha_{k,m}\sim \mathcal{CN}(0,\sigma_{\alpha}^{2})$ denotes the complex reflection coefficient of the $k$-th target at the $m$-th subcarrier.
$\bm \theta_{\text{s}}=[\theta_{\text{s},1}, \cdots, \theta_{\text{s},K}]$ and $\bm \phi_{\text{s}}=[\phi_{\text{s},1}, \cdots, \phi_{\text{s},K}]$ denote the elevation and the azimuth angles of the $K$ targets respectively. 
$\bm A_{t,m}=[\bm a_{t}(\theta_{\text{s},1}, \phi_{\text{s},1},f_{m}),\cdots,\bm a_{t}(\theta_{\text{s},K}, \phi_{\text{s},K},f_{m})]$, and $\Sigma_{m}=\text{diag}(\alpha_{1,m}, \cdots,\alpha_{K,m})$.
Then the echo at the $m$-th subcarrier at BS can be written as
\vspace{-0.2cm}
\begin{equation}
    \bm y_{\text{s},m}= \sqrt{p_{m}}\bm A_{r,m}\bm\Sigma_{m} \bm A_{t,m}^{T} \bm F_{\text{TD},m} \bm f_{\text{PS}} s_{m}+\bm n_{\text{s},m},
\vspace{-0.1cm}
\end{equation}
with $\bm n_{\text{s},m}\sim\mathcal{CN}(\bm 0, \sigma_{\text{s}}^{2}\bm I_{N_{r}})$ being the ISAC receiver noise. 
We consider whitening the received noise to enforce adherence to a complex Gaussian distribution.
Accordingly, the Fisher information matrix with respect to $\bm \theta$ and $\bm \phi$ is expressed as
\vspace{-0.2cm}
\begin{equation}
\bm J_{m}=\frac{2}{\sigma_{\text{s}}^{2}}\text{Re}
\left[ 
\begin{matrix}
\bm J_{m}(\bm \theta_{\text{s}}, \bm \theta_{\text{s}}) &  \bm J_{m}(\bm \theta_{\text{s}}, \bm \phi_{\text{s}})   \\
\bm J_{m}^{\text{H}}(\bm \theta_{\text{s}}, \bm \phi_{\text{s}}) & \bm J_{m}(\bm \phi_{\text{s}}, \bm \phi_{\text{s}}) \\
\end{matrix}
\right],
\vspace{-0.2cm}
\end{equation}
where
\vspace{-0.2cm}
\begin{equation}
\begin{aligned}
    \bm J_{m}(\bm \theta_{\text{s}}, \bm \phi_{\text{s}})&=(\dot{\bm A}_{r,\theta,m}^{\text{H}} \dot{\bm A}_{r,\phi,m})\odot (\bm \Sigma_{m}^{*}\bm A_{t,m}^{\text{H}}\bm R_{x,m}^{*}\bm A_{t,m}\bm \Sigma_{m} ) \\
    &+ (\dot{\bm A}_{r,\theta,m}^{\text{H}} \bm A_{r,m})\odot (\bm \Sigma_{m}^{*}\bm A_{t,m}^{\text{H}}\bm R_{x,m}^{*}\dot{\bm A}_{t,\phi,m}\bm \Sigma_{m} ) \\
    &+(\bm A_{r,m}^{\text{H}} \dot{\bm A}_{r,\phi,m})\odot (\bm \Sigma_{m}^{*} \dot{\bm A}_{t,\theta,m}^{\text{H}}\bm R_{x,m}^{*}\bm A_{t,m}\bm \Sigma_{m} ) \\
    &+(\bm A_{r,m}^{\text{H}} \bm A_{r,m})\odot (\bm \Sigma_{m}^{*}\dot{\bm A}_{t,\theta,m}^{\text{H}}\bm R_{x,m}^{*}\dot{\bm A}_{t,\phi,m}\bm \Sigma_{m} ),  \nonumber
\end{aligned}
\vspace{-0.2cm}
\end{equation}
with $\dot{\bm A}_{\text{i},\theta,m}=\left[\frac{\partial \bm a_{\text{i}}(\theta_{\text{s},1},\phi_{\text{s},1},f_{m})}{\partial \theta_{\text{s},1}},\cdots, \frac{\partial \bm a_{\text{i}}(\theta_{\text{s},K},\phi_{\text{s},K},f_{m})}{\partial \theta_{\text{s},K}}\right ]$ and $\dot{\bm A}_{\text{i},\phi,m}=\left[\frac{\partial \bm a_{\text{i}}(\theta_{\text{s},1},\phi_{\text{s},1},f_{m})}{\partial \phi_{\text{s},1}},\cdots, \frac{\partial \bm a_{\text{i}}(\theta_{\text{s},K},\phi_{\text{s},K},f_{m})}{\partial \phi_{\text{s},K}}\right ]$ ($\text{i}=\{t,r\}$), and $\bm R_{x,m}=\mathbb{E}\left[\bm x_{m}\bm x_{m}^{\text{H}}\right]=p_{m}\bm F_{\text{TD},m}\bm f_{\text{PS}}\bm f_{\text{PS}}^{\text{H}}\bm F_{\text{TD},m}^{\text{H}}$ denotes the covariance matrix of $\bm x_{m}$.
Therefore the sensing performance indicator of this system, Cramér-Rao bound (CRB) associated with angle estimation, is calculated as
\vspace{-0.25cm}
\begin{equation}
    \text{CRB}(\bm \theta_{\text{s}}, \bm \phi_{\text{s}})=\text{Tr}\left[\left(\sum_{m=1}^{M}\bm J_{m}\right)^{-1}\right],
    \vspace{-0.1cm}
\end{equation}
which is determined by both the analog precoder and the distribution of the targets. The prior angle information, obtained at the beginning of each frame \cite{frame_ISAC}, is used to optimize the CRB for the duration of data transmission.

\vspace{-0.1cm}
\subsection{Dual-Functional Gain: ISAC Performance Indicator}
\vspace{-0.0cm}
To comprehensively evaluate the performance of both functionalities, we define the dual-functional gain as the overall metric. Given the communication and sensing channels $\bm h_{\text{c}}=\{\bm h_{\text{c},m}\}_{m=1}^{M}$ and $\bm G=\{\bm G_{m}\}_{m=1}^{M}$, we characterize the rate-CRB region corresponding to all rate-CRB pairs that can be simultaneously achieved under the total power $P_{t}$, i.e., 
\vspace{-0.2cm}
\begin{equation}
    \mathcal{C}^{\text{ISAC}}(P_{t},\bm h_{\text{c}}, \bm G)\triangleq \bigcup_{\bm T,\bm f_{\text{PS}},\bm p}\{(x, y)\vert x\leqslant R_{\text{c}}, y\geqslant\text{CRB}(\bm \theta_{\text{s}}, \bm \phi_{\text{s}}) \}. 
    \vspace{-0.1cm}
\end{equation}
Denote $(R_{\text{sen}}, \text{CRB}_{\text{min}})$ as the point achieved by the sensing-dedicated scheme and $(R_{\text{max}}, \text{CRB}_{\text{com}})$ by the communication-dedicated scheme. Then $\mathcal{C}^{\text{ortho}}(P_{t},\!\bm h_{c},\!\bm G)$ denotes the achievable region under non-ISAC orthogonal operations, bounded by the line segment between $(R_{\text{sen}}, \text{CRB}_{\text{min}})$ and $(R_{\text{max}}, \text{CRB}_{\text{com}})$.
Then the dual-functional gain can be measured as 
\vspace{-0.2cm}
\begin{equation}
\varrho=\frac{\int_{(x,y)\in \mathcal{C}^{\text{ISAC}}\backslash\mathcal{C}^{\text{ortho}}} dxdy}{(R_{\text{max}}-R_{\text{sen}})(\text{CRB}_{\text{com}}-\text{CRB}_{\text{min}})/2}\in [0,1),
\vspace{-0.2cm}
\end{equation}
representing the expansion of the Pareto boundary of $\mathcal{C}^{\text{ISAC}}$ compared to that of $\mathcal{C}^{\text{ortho}}$. $\varrho=0$ denotes the orthogonal operation and $\varrho\rightarrow 1$ means both $R_{\text{max}}$ and $\text{CRB}_{\text{min}}$ are simultaneously approached.
Our objective is to jointly optimize TTDs, PSs and power allocation across subcarriers for maximizing the dual-functional gain.

\section{SA-Opt: A Near-Optimal Benchmark}

In this section, we propose SA-Opt, a squint-aware analog precoding optimization scheme based on existing sub-THz hardware as a benchmark. First, we define the C-S channel correlation to quantify the relationship between the two functionalities and demonstrate its positive impact on dual-functional gain. Based on this, we optimize the TTDs to maximize the correlation between equivalent communication and sensing channels, paving the way for enhanced dual-functional gain. Finally, we optimize the PSs and transmitting power allocation to approach the Pareto boundary.

\vspace{-0.1cm}
\subsection{Definition of C-S Channel Correlation}

We evaluate the C-S channel correlation by examining the beamspace power distribution across subcarriers. The communication and sensing channels are first transformed into the beamspace through Discrete Fourier Transform (DFT) as
\vspace{-0.2cm}
\begin{equation}
    \bm h^{b}_{\text{c},m}=\bm D_{t}^{\text{H}}\bm h_{\text{c},m},~~\bm h^{b}_{\text{s},m}=\text{diag}(\bm D_{r}^{\text{H}}\bm G_{m}\bm D_{t}),
    \vspace{-0.1cm}
\end{equation}
where $\bm D_{t}\in \mathbb{C}^{N_{t}\times N_{D}}$ and $\bm D_{r}\in \mathbb{C}^{N_{r}\times N_{D}}$ are the dictionary uniformly sampled in beamspace \cite{shijian_MI_max}. Note that $\bm G_{m}$ is composed of the steering vectors at both ends. Therefore, we transform the transmitter and receiver separately into the beamspace.
Then the beamspace channels across all subcarriers are summed and normalized to represent the discrete spatial distribution of sensing and communication tasks as
\vspace{-0.2cm}
\begin{equation}
    \widehat{\bm h}^{b}_{\text{c}}= \frac{\sum_{m=1}^{M}\vert \bm h^{b}_{\text{c},m} \vert}{\| \sum_{m=1}^{M}\vert \bm h^{b}_{\text{c},m} \vert \|_{1}},~~\widehat{\bm h}^{b}_{\text{s}}=\frac{\sum_{m=1}^{M}\vert \bm h^{b}_{\text{s},m} \vert}{\| \sum_{m=1}^{M}\vert \bm h^{b}_{\text{s},m} \vert \|_{1}}.
    \vspace{-0.1cm}
\end{equation}
With the beamspace channels $\widehat{\bm h}^{b}_{\text{c}}$ and $\widehat{\bm h}^{b}_{\text{s}}$ defined, the communication-sensing (C-S) channel correlation is quantified as follows:
\vspace{-0.15cm}
\begin{equation}
    \text{Cor}(\bm h_{\text{c}}, \bm G)=\frac{1}{\text{KL}( \widehat{\bm h}^{b}_{\text{c}}, \widehat{\bm h}^{b}_{\text{s}}  )},
    \vspace{-0.1cm}
\end{equation}
where $\text{KL}(\bm p, \bm q)=\sum_{n=1}^{N} p[n]\ln \frac{p[n]}{q[n]}$ denotes the Kullback-Leibler (K-L) divergence between distributions $\bm p$ and $\bm q$.

\vspace{-0.2cm}
\begin{proposition}
Denote $R^{\star}(\bm h_{\text{c}}, \bm G)$ and $\text{CRB}^{\star}(\bm h_{\text{c}}, \bm G)$ as a rate-CRB pair on the Pareto boundary of $\mathcal{C}^{\text{ISAC}}(\bm h_{c}, \bm G)$. Under the same transmitting power, both $R^{\star}(\bm h_{\text{c}}, \bm G)$ and $(\text{CRB}^{\star}(\bm h_{\text{c}}, \bm G))^{-1}$ increase with higher $\text{Cor}(\bm h_{\text{c}}, \bm G)$. In other words, the dual-functional gain improves with higher C-S channel correlation.
\end{proposition}
\vspace{-0.2cm}
\begin{proof}
See Appendix A.
\end{proof}
\vspace{-0.1cm}

\subsection{C-S Channel Correlation Maximization through TTDs}
\vspace{-0.0cm}
Note that TTDs contribute to adjusting the equivalent wideband channels, i.e., $\widetilde{\bm h}_{\text{c},m}=\bm F_{\text{TD,m}}^{\text{H}}\bm h_{\text{c},m}$ and $\widetilde{\bm G}_{m}=\bm G_{m}\bm F_{\text{TD,m}}$. We first design TTDs to maximize the C-S channel correlation by solving
\vspace{-0.2cm}
\begin{align}
    \underset{ {\bm T }}{\max}&~ \text{Cor}(\widetilde{\bm h}_{\text{c}}, \widetilde{\bm G}) \label{Problem1}\\
    \text{s.t.} &~ \bm F_{\text{TD},m}=\text{diag}\left[\text{vec}\left( e^{j2\pi f_{m}(\bm T \otimes \bm 1^{L_{h}\times L_{v}}) } \right) \right ], \tag{17a} \label{Problem1_a}\\
    &~  \bm T[q_{h}, q_{v}]\in \mathcal{T}. \tag{17b}\label{Problem1_b}
    \vspace{-0.3cm}
\end{align}
Since both the objective and the constraints are complicated, we conduct an iterative exhaustive search on $\bm T[q_{h}, q_{v}]$ in the feasible region $\mathcal{T}$ in an element-wise manner.

\vspace{-0.1cm}
\subsection{Alternating Optimization of $\bm f_{\text{PS}}$ and $\bm p$}
\vspace{-0.0cm}
After TTDs are fixed, the PSs and the power allocation across subcarriers are optimized to improve ISAC performance under the adjusted equivalent channels. $\bm f_{\text{PS}}$ and $\bm p$ are optimized in an alternating manner. 

\setlength{\parindent}{0pt}
\textbf{Updating $\bm f_{\text{PS}}$:}
\setlength{\parindent}{10pt}

We first optimize $\bm f_{\text{PS}}$ with fixed $\bm p$.
Since maximizing the array gain in sub-THz systems is crucial for enhancing both communication and sening performance \cite{crb_opt}, we design $\bm f_{\text{PS}}$ to maximize the weighted sum of the array gains at both the user and the echo reception. The problem is formulated as
\vspace{-0.2cm}
\begin{align}
    \underset{ {\bm f_{\text{PS}}} }{\max}\!&~ \sum_{m=1}^{M}p_{m}\left[ \| \widetilde{\bm h}_{\text{c},m}^{\text{H}}\bm f_{\text{PS}}\|_{2}^{2} + \frac{\eta}{N_{r}}\| \widetilde{\bm G}_{\text{s},m}\bm f_{\text{PS}}\|_{2}^{2} \right] \label{Problem2} \\
    \text{s.t.} &~ \left| \bm f_{\text{PS}}[n] \right| =1,~n\in [1, N_{t}], \tag{18a} \label{Problem2_a}
    \vspace{-0.2cm}
\end{align}
where $\eta$ is the weighting coefficient to balance array gain for sensing and communications. Using Cauchy-Schwarz inequality for relaxation, a closed-form solution is
\vspace{-0.2cm}
\begin{equation}
    \bm f_{\text{PS}}=\text{exp}\left[j\angle \left(\sum_{m=1}^{M} \sqrt{p_{m}} (\widetilde{\bm h}_{\text{c},m} + \sum_{r=1}^{N_{r}}\frac{\eta}{N_{r}}\widetilde{\bm G}^{\text{H}}_{m}[r,:] ) \right)\right]. \tag{19}
    \label{equ:PS}
    \vspace{-0.0cm}
\end{equation}

\vspace{-0.1cm}
\setlength{\parindent}{0pt}
\textbf{Updating $\bm p$:}
\setlength{\parindent}{10pt}

With $\bm f_{\text{PS}}$ updated, the power allocation across subcarriers is optimized to minimize the CRB of targets' angles estimation while ensuring compliance with the user's communication rate requirements, by solving
\vspace{-0.25cm}
\begin{align}
    \underset{ {\bm p}}{\min}&~ \text{CRB}(\bm \theta_{\text{s}}, \bm \phi_{\text{s}}) \label{Problem4} \tag{20}\\
    \text{s.t.} &~ \vert \overline{h}_{\text{c},m}\vert^{2}p_{m} \geqslant \Gamma, \tag{20a} \label{Problem4_a}\\
    &~ \| \bm p \|_{1}\leqslant \frac{P_{t}}{N_{t}},  \tag{20b} \label{Problem4_b}
    \vspace{-0.35cm}
\end{align}
where $\overline{h}_{\text{c},m}=\widetilde{\bm h}_{\text{c},m}^{\text{H}}\bm f_{\text{PS}}$ is the equivalent digital channel, and $\Gamma$ is the signal-to-noise ratio (SNR) threshold at the user. 
Note that the objective can be relaxed as $\text{Tr}\left[(\sum_{m=1}^{M} p_{m} \bm J_{m} )^{-1} \right ] \leqslant \frac{1}{M^{2}}\sum_{m=1}^{M}\frac{1}{p_{m}}  \text{Tr}(\bm J_{m}^{-1})$ via Jensen's inequality, giving rise to
\vspace{-0.25cm}
\begin{align}
    \underset{ {\bm p}}{\min}&~ \text{Tr}\left(\bm Q (\text{diag}(\bm p))^{-1}\right)
    \label{Problem5} \tag{21}\\
    \text{s.t.} &~ (\ref{Problem4_a}),(\ref{Problem4_b}), \nonumber
    \vspace{-0.35cm}
\end{align}
where $\bm Q=\text{diag}([ \text{Tr}(\bm J_{1}^{-1}), \cdots, \text{Tr}(\bm J_{M}^{-1}) ])$.
This problem is a convex programming and can be solved by the CVX toolbox. The pseudo code of the SA-Opt approach is illustrated in Algorithm~\ref{Alg_1}.

\begin{figure*}[t]
  \vspace{-0.0cm}
  \setlength{\abovecaptionskip}{-0cm} 
  \setlength{\belowcaptionskip}{0.1cm} 
  \centering
  \includegraphics[width=0.94\linewidth]{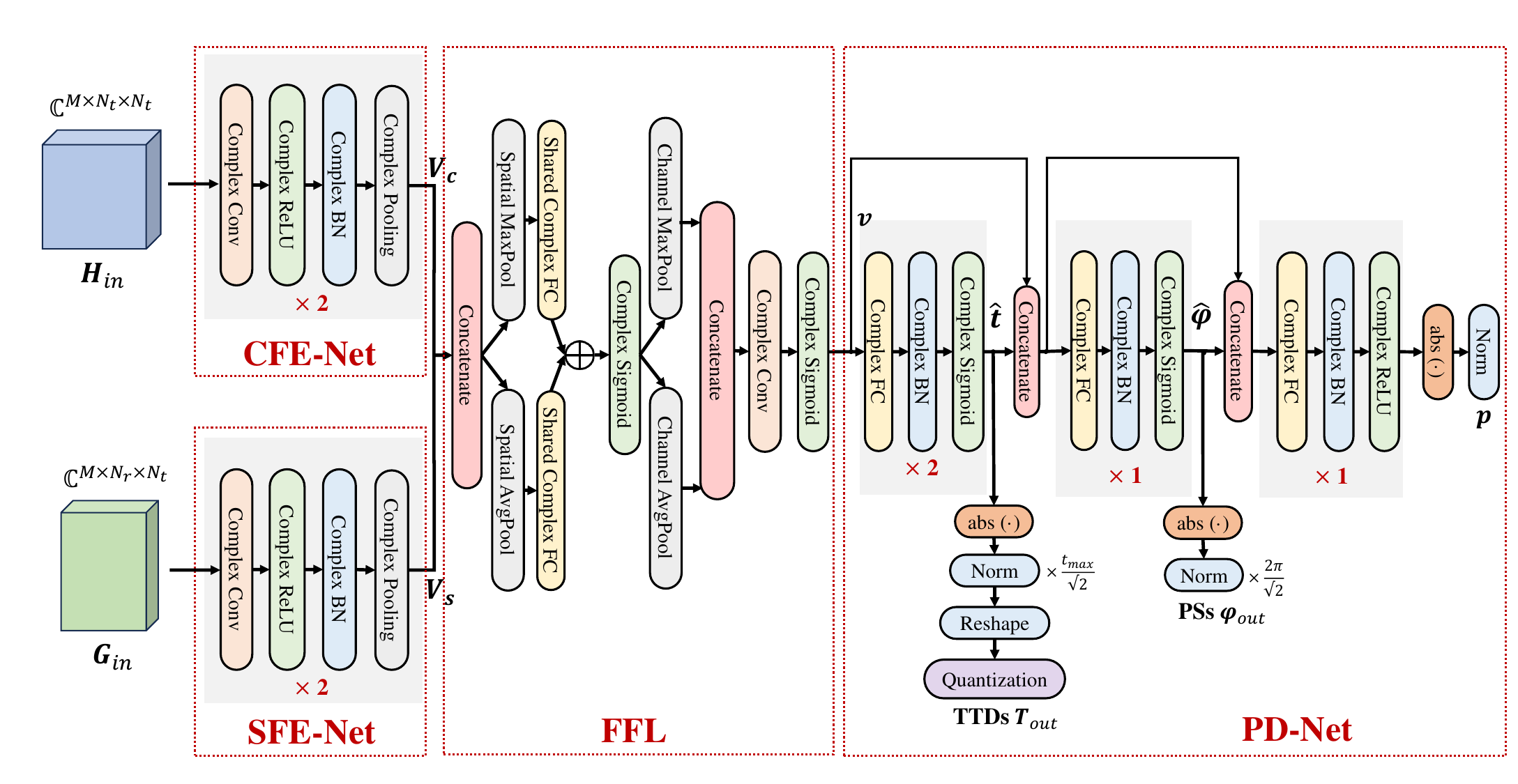}
  \captionsetup{font=small}
  \caption{The architecture of CSP-Net. The CSP-Net is complex-valued and consists of four parts: Communication channel feature extraction network (CFE-Net), sensing channel feature extraction network (SFE-Net), feature fusion layer (FFL) and parameter design network (PD-Net).}
  \label{fig:network architecture}
  \vspace{-0.55cm}
\end{figure*}

\subsection{Complexity Analysis of SA-Opt}

The computational complexity of element-wise update of $\bm T$ is $\mathcal{O}(N_{\text{iter}}Q_{t}2^{B_{t}}(N_{t}^{3}N_{r}^{3}\!+\!2N_{t}^{2}N_{r}+2N_{t}^{2}))$, with $N_{\text{iter}}$ being the number of  updates for $\bm T$.
The complexity of optimizing $\bm f_{\text{PS}}$ and $\bm p$ with an accuracy $\epsilon$ is $\mathcal{O}(N_{\text{AO}}(M((2K)^{3}\!+\!12N_{t}^{2}K^{4})\!+\!M^{3}\log{1/\epsilon}))$, with $N_{\text{AO}}$ being the number of alternating iterations for $\bm f_{\text{PS}}$ and $\bm p$. Thus, the overall computational complexity of SA-Opt scales with the cube of $N_{t}$ and grows exponentially with the TTD resolution $B_{t}$, depending on the number of iterations.

\vspace{-0.1cm}
\begin{algorithm}[h]
  \caption{Squint-Aware Optimization-Based Analog Precoder Design Algorithm for ISAC (SA-Opt)}
  \textbf{Input}:$\{\bm h_{\text{c},m}\}_{m=1}^{M}$, $\{\bm G_{m}\}_{m=1}^{M}$, $M$, $N_{t}$, $N_{r}$, $Q_{h}$, $Q_{v}$, $P_{t}$, $\eta$, $\Gamma$, $\sigma_{\text{c}}$, $\sigma_{\text{s}}$, $N_{\text{iter}}$, $N_{\text{AO}}$;\\
  \textbf{Output}:$\bm T, \bm f_{\text{RF}}$ and $\bm p$; \\
  \textbf{Steps}:
  \begin{algorithmic}[1]
  \State Initialize $\bm T$ and $\bm f_{\text{PS}}$ randomly, initialize $\bm p=\frac{P_{t}}{N_{t}M}\bm 1$;
  \For {$n=1$ to $N_{\text{iter}}$}
  \For{$q_{h}=1$ to $Q_{h}$}
  \For{$q_{v}=1$ to $Q_{v}$}
  \State Set $\bm T[q_{h},q_{v}]\!=\!\mathop{\arg\!\max}\limits_{t\in \mathcal{T}} \text{Cor}(\bm h_{\text{c}},\!\bm G_{\text{s}}; \bm T[q_{h},q_{v}]\!=\!t)$;
  \EndFor
  \EndFor
  \EndFor
  \For{$n=1$ to $N_{\text{AO}}$}
  \State Update $\bm f_{\text{PS}}$ as (\ref{equ:PS});
  \State Update $\bm p$ by solving (\ref{Problem5});
  \EndFor
  \end{algorithmic}
  \label{Alg_1}
\end{algorithm}
\vspace{-0.1cm}

\section{CSP-Net: Low-Complexity Alternative}
\vspace{-0.1cm}

The proposed SA-Opt exhibits high computational complexity due to its search and iterative processes. This may result in excessive latency in the analog precoding design, preventing it from meeting the requirements of future mobile networks.
Therefore, in this section, we present a low-complexity alternative based on unsupervised learning, termed as the complex-valued squint-aware analog precoding network (CSP-Net).

\vspace{-0.15cm}
\subsection{Complex-Valued Network Structure}
\vspace{-0.05cm}
The commonly used real-valued networks in \cite{learning_ISAC_Elbir, learning_ISAC_precoding} may fail to capture the phase information between the real and imaginary components of the complex-valued channel data. Therefore, we employ a complex-valued neural network architecture specifically designed for complex channel inputs.
This network mainly includes complex-valued convolutional layers, complex-valued fully-connected (FC) layers, complex-valued batch normalization (BN) layers, and complex-valued activation functions, as detailed in \cite{complex_network}.

The input contains the communication and sensing channels $\{\bm h_{\text{c},m}, \bm G_{m}\}_{m=1}^{M}$. 
We pre-process the channel data as $\overline{\bm H}_{m} = \bm h_{\text{c},m}\bm h_{\text{c},m}^{\text{H}}$ and $\overline{\bm G}_{m} = \bm G_{m}$ respectively.
Then the total data are represented as $\overline{\bm H}=\{\overline{\bm H}_{m}\}_{m=1}^{M}\in \mathbb{C}^{M\times N_{t}\times N_{t}}$ and $\overline{\bm G}=\{\overline{\bm G}_{m}\}_{m=1}^{M}\in \mathbb{C}^{M\times N_{r}\times N_{t}}$.
Following this, we perform power normalization before inputting them into the neural network.

\subsection{Layer Description}

The proposed CSP-Net consists of four parts: Communication channel feature extraction network (CFE-Net), sensing channel feature extraction network (SFE-Net), feature fusion layer (FFL) and parameter design network (PD-Net).
The detailed architecture design of CSP-Net is illustrated in Fig.~\ref{fig:network architecture}.

\vspace{-0.15cm}
\subsubsection{CFE-Net and SFE-Net}

The features from the communication and sensing channels are extracted respectively by CFE-Net and SFE-Net. For communication channel data $\bm H_{in}$,  two layers of complex convolutional blocks are employed, each of which includes one complex convolutional layer, one complex ReLU, one complex batch normalization (BN) and one complex pooling. Similarly, another branch of convolutional blocks is employed for $\bm G_{in}$ in SFE-Net.
Parameterized by $\phi_{\text{CFE}}$ and $\phi_{\text{SFE}}$, the outputs are represented as $\bm V_{\text{c}}=\phi_{\text{CFE}}(\bm H_{in})$ and $\bm V_{\text{s}}=\phi_{\text{SFE}}(\bm G_{in})$ respectively.

\vspace{-0.15cm}
\subsubsection{Feature Fusion Layer}
To further enhance the features from $\bm H_{in}$ and $\bm G_{in}$ and extract the correlation between them, we tailor the convolutional block attention module (CBAM) similar to \cite{CBAM} on $[\bm V_{\text{c}}, \bm V_{\text{s}}]$ with the complex-valued network, including a channel-based attention and a spatial-based attention. Then the output is flattened as $\bm v$ for the following precoding design.

\vspace{-0.15cm}
\subsubsection{PD-Net}
The values of TTDs, PSs and the power allocation are sequentially predicted through PD-Net.
Two complex FC layers are first employed to predict the value of TTDs, with the complex sigmoid function as the activation function. 
However, the output, $\hat{\bm t}\in \mathbb{C}^{Q_{t}}$, fails to satisfy the finite-value finite-resolution constraint. Therefore the post-processing is conducted to project $\hat{\bm t}$ into the feasible region as $\bm T_{out}=\mathcal{Q}(\text{reshape}(\frac{t_{\text{max}}}{\sqrt{2}}\vert\hat{\bm t}\vert))$, where $\mathcal{Q}(\bm T[q_{h}, q_{v}])=\mathop{\arg\min}\limits_{t\in \mathcal{T}}\|t- \bm T[q_{h}, q_{v}]\|_{2}^{2}$.
Concatenating $\bm v$ and $\hat{\bm t}$ as the input of the following network, we proceed to predict the PSs through one complex FC layer, and the output is expressed as $\hat{\bm \varphi}\in \mathbb{C}^{N_{t}}$. Then PSs are calculated as $\bm \varphi_{out}=\frac{2\pi}{\sqrt{2}}\vert \hat{\bm \varphi}\vert$.
Since the design of power allocation is determined by both the channels and the configuration of analog parts, we concatenate $\bm v$, $\hat{\bm t}$ and $\hat{\bm \varphi}$ to predict $\bm p$ using another complex FC layer.

The overall parameters in the proposed CSP-Net are $\phi_{\text{CSP}}=\{\phi_{\text{CFE}}, \phi_{\text{SFE}}, \phi_{\text{FFL}}, \phi_{\text{PD}} \}$.

\vspace{-0.15cm}
\subsection{Loss Function}
We employ an unsupervised learning-based method for the model training. To achieve the goal of maximizing ISAC utility with the awareness of the C-S channel correlation, we customize the loss function for one batch as
\vspace{-0.15cm}
\begin{equation}
    L(\phi_{\text{CSP}})=-\frac{1}{N_{b}}\sum_{i=1}^{N_{b}}\frac{\text{Cor}(\bm h_{\text{c}}^{(i)}, \bm G^{(i)};\bm T^{(i)})}{\text{Cor}^{\star}(\bm h_{\text{c}}^{(i)}, \bm G^{(i)};\bm T^{(i)})}\left[\frac{R_{\text{c}}^{(i)}}{R_{\text{max}}^{(i)}}+\frac{\text{CRB}_{\text{min}}^{(i)}}{\text{CRB}^{(i)}}\right ],
    \label{equ:loss}
    \tag{22}
    \vspace{-0.1cm}
\end{equation}
with $N_{b}$ denoting the batch size.
In offline training stage, CSP-Net is updated with this loss function by back propagation in randomly-sampled batches. In the deployment, the channel data are input to the well-trained CSP-Net, and the values of TTDs, PSs and power allocation are obtained sequentially.

\begin{figure*}[t]
\vspace{-0.0cm}
\centering
\subfigure[Loss function.]{\label{fig:subfig:a}
\includegraphics[width=0.24\linewidth]{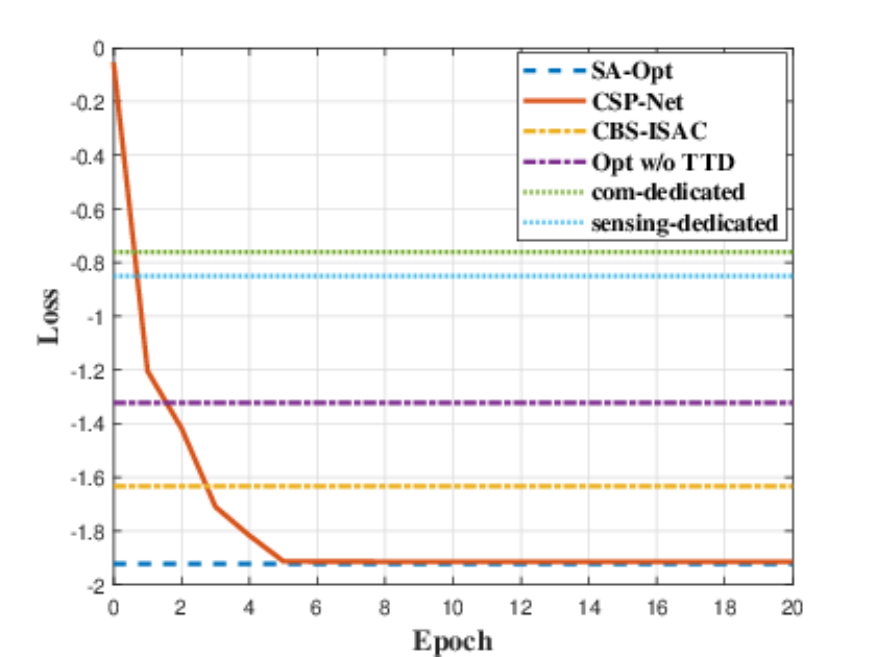}}
\hspace{-0.01\linewidth}
\subfigure[Normalized correlation.]{\label{figs}
\includegraphics[width=0.24\linewidth]{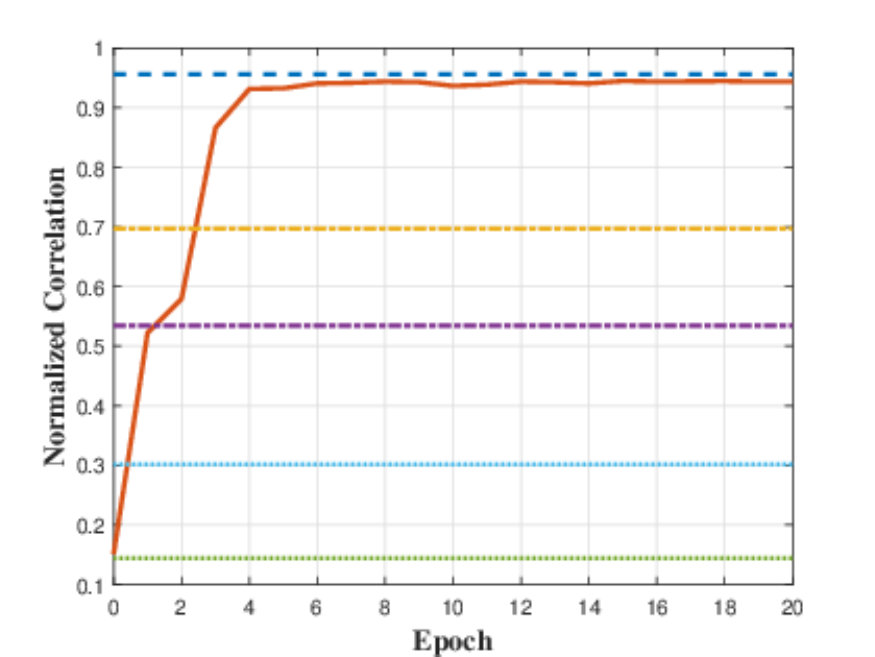}}
\hspace{-0.01\linewidth}
\subfigure[Rate.]{\label{fig:subfig:a}
\includegraphics[width=0.24\linewidth]{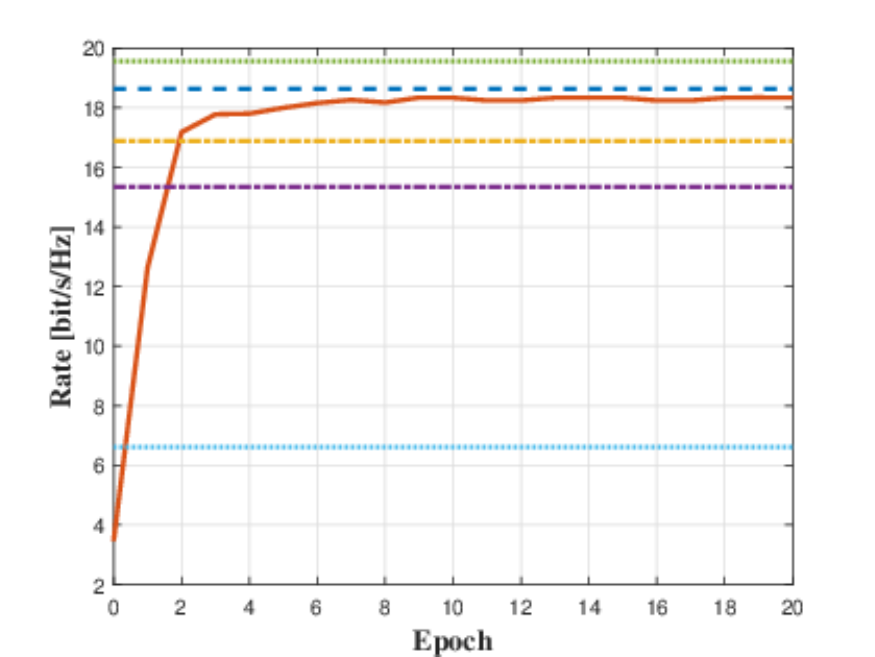}}
\hspace{-0.01\linewidth}
\subfigure[CRB.]{\label{fig:subfig:b}
\includegraphics[width=0.24\linewidth]{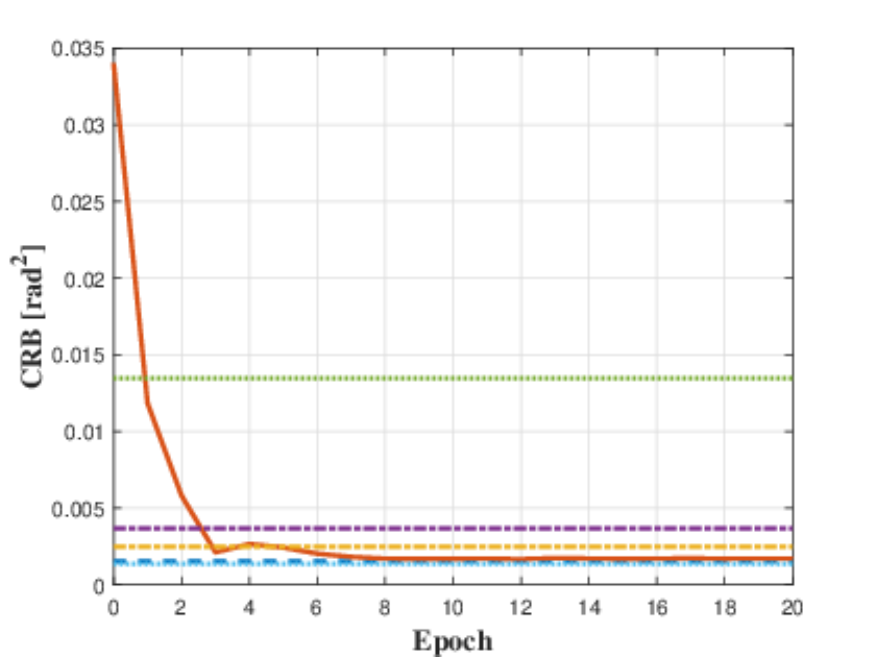}}
\label{fig:subfig}
\vspace{-0.3cm}
\captionsetup{font=small}
\caption{Loss and performance versus training epochs. The proposed CSP-Net converges after 10 epochs. Throughout the training process, both communication and sensing performances improve as the C-S channel correlation increases.}
\label{fig:training}
\vspace{-0.45cm}
\end{figure*}

\section{Existing Method: Controlled Beam Squint}

To verify the effectiveness of proposed squint-aware methods, we further develop a benchmark scheme that directly controls beam squint for ISAC without correlation adjustment.

Denote $\{\bm F_{m}^{\star}\}_{m=1}^{M}$ as the ideal frequency-dependent analog precoding matrices in this hardware structure, making the beams at all subcarriers point to the desired directions $\{(\theta_{m}^{\star}, \phi_{m}^{\star})\}_{m=1}^{M}$ to guarantee the coverage of both targets and users.
Note that the design of $\{(\theta_{m}^{\star}, \phi_{m}^{\star})\}_{m=1}^{M}$ is equivalent to the shortest path planning problem in the two-dimensional angular plane, and will not be elaborated further here.
To achieve the ideal array gain in these directions, the precoder at $f_{m}$ is set as the steering vector towards ($\theta_{m}^{\star}, \phi_{m}^{\star}$), i.e., $\bm F_{m}^{\star}=\bm a_{t}(\theta_{m}^{\star},\phi_{m}^{\star}, f_{m})$.
Subsequently, $\bm T$ and $\bm f_{\text{PS}}$ are jointly designed to approximate $\{\bm F_{m}^{\star}\}_{m=1}^{M}$ by solving
\vspace{-0.25cm}
\begin{align}
    \underset{ {\bm T, \bm \varphi}}{\min}&~ \frac{1}{M}\sum_{m=1}^{M}\| \bm F_{\text{TD},m}\bm f_{\text{PS}}-\bm F_{m}^{\star}\|_{\text{F}}^{2} \label{Problem11} \tag{23}\\
    \text{s.t.} &~  (\ref{Problem1_a}), (\ref{Problem1_b}), (\ref{Problem2_a}). \nonumber
    \vspace{-0.35cm}
\end{align}
According to Lemma 1 in \cite{DPP_constrained_TTD}, this problem can be equivalently transformed into a convex one as 
\vspace{-0.2cm}
\begin{align}
    \underset{ {\bm \varphi, \bm T}}{\min}&  \frac{1}{M}\sum_{m=1}^{M}\sum_{q_{h}=1}^{Q_{h}}\sum_{q_{v}=1}^{Q_{v}}\sum_{l_{h}=1}^{L_{h}}\sum_{l_{v}=1}^{L_{v}}\vert \bm \varphi[(q_{h}-1)L_{h}+l_{h}, (q_{v}-1)L_{v}+l_{v}] \nonumber \\
    &-2f_{m}\bm T[q_{h}, q_{v}] +\frac{f_m}{f_{\text{c}}}[\sin{\theta_m}\sin{\phi_m}((q_{h}-1)L_{h}+l_{h}-1) \nonumber \\
    &+ \cos{\theta_m}((q_{v}-1)L_{v}+l_{v}-1)]  \vert^2 \label{Problem12} \tag{24}\\
    \text{s.t.} &~ 0\leqslant  \bm T[q_{h},q_{v}] \leqslant t_{\text{max}}, \tag{24a}\label{Problem12_a} \\
    &~ 0\leqslant \bm \varphi[n] < 2\pi, \tag{24b}\label{Problem12_b} 
    \vspace{-0.3cm}
\end{align}
and solved by the off-the-shelf toolbox. $\bm T$ is then quantized to meet the finite-resolution constraint.
The remaining task is to optimize $\bm p$ with $\bm T$ and $\bm \varphi$ fixed, as detailed in Section 3.3.

\section{Experimental Results}

In this section, simulation results are presented to evaluate the performance of the proposed schemes.

\begin{figure*}[t]
\vspace{-0.0cm}
\centering
\subfigure[Rate versus SNR.]{\label{fig:rate_snr}
\includegraphics[width=0.45\linewidth]{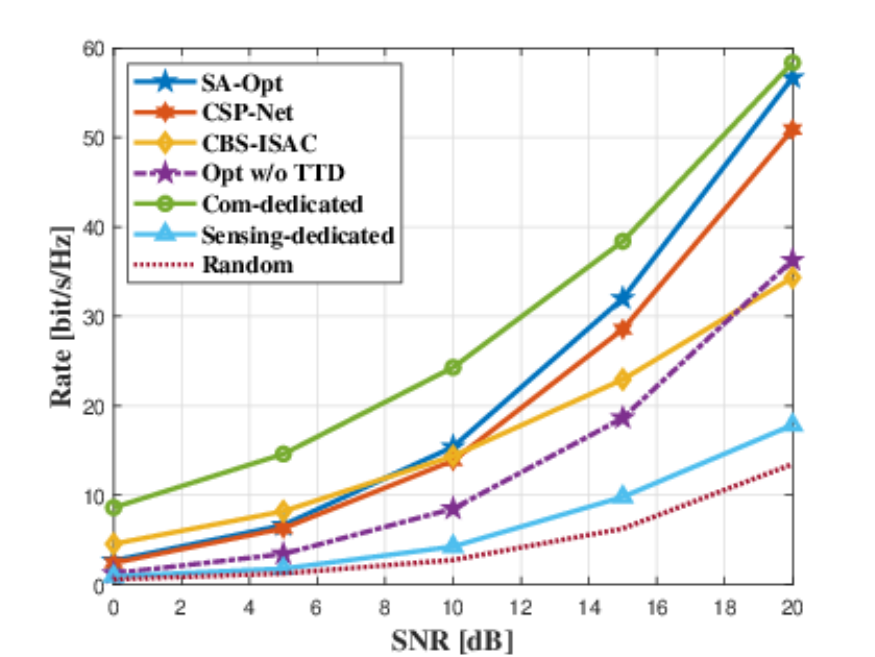}}
\hspace{0.000001\linewidth}
\subfigure[CRB versus SNR.]{\label{fig:crb_snr}
\includegraphics[width=0.45\linewidth]{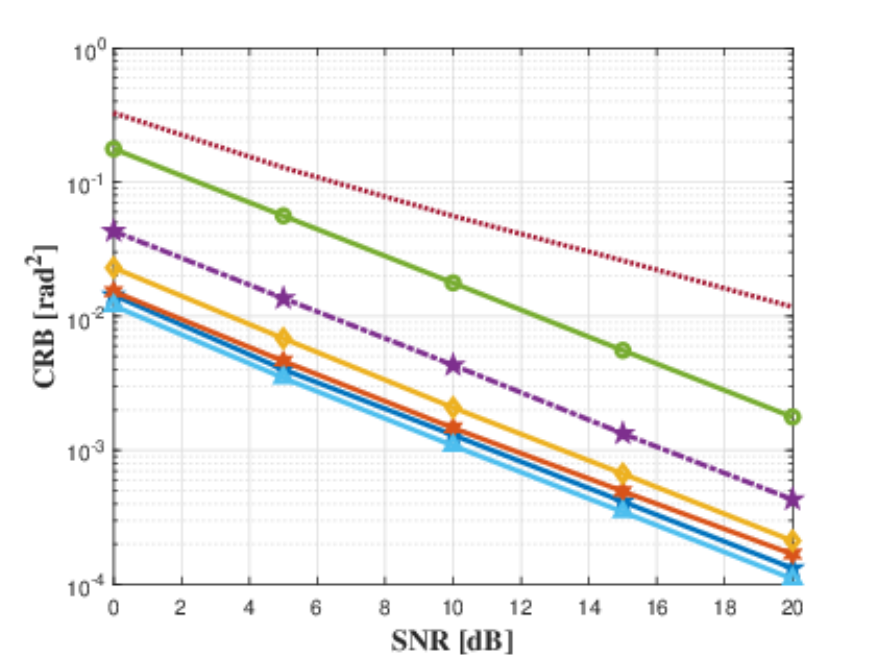}}
\hspace{0.000001\linewidth}
\subfigure[Rate versus MSIA.]{\label{fig:rate_msia}
\includegraphics[width=0.45\linewidth]{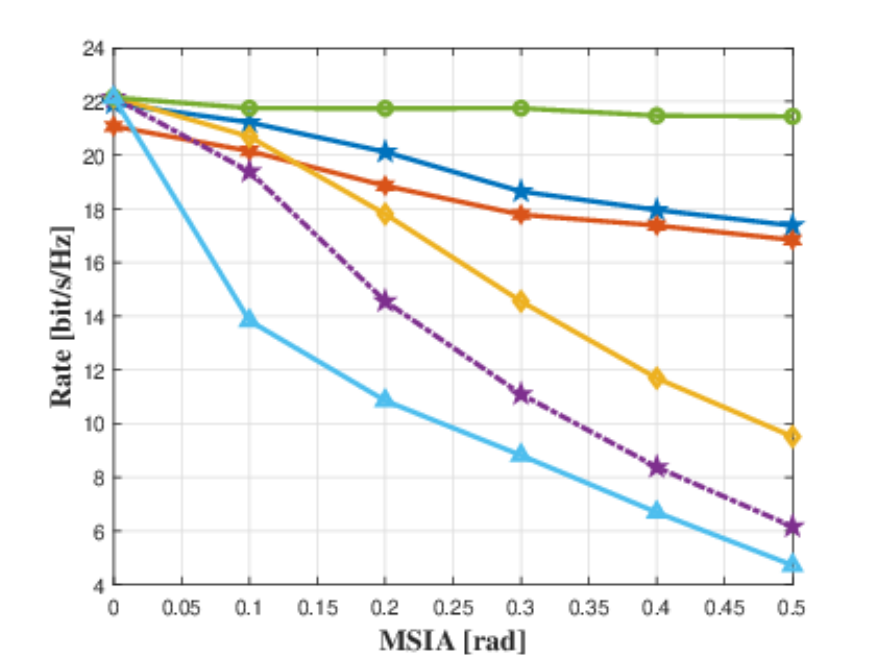}}
\hspace{0.000001\linewidth}
\subfigure[CRB versus MSIA.]{\label{fig:crb_msia}
\includegraphics[width=0.45\linewidth]{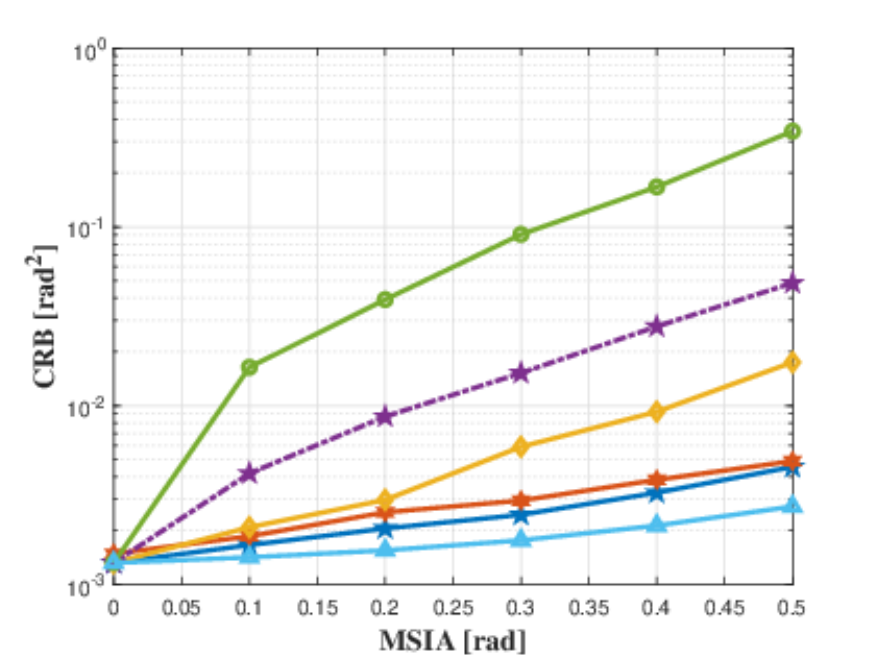}}
\vspace{-0.3cm}
\captionsetup{font=small}
\caption{ISAC performance comparison. The proposed squint-aware analog precoder designs outperform existing methods through proactive correlation adjustment, improving both the communication rate and the sensing accuracy. These methods offer better adaptation to spatial distribution, maintaining performance even as the MSIA increases.}
\label{fig:SNR}
\vspace{-0.6cm}
\end{figure*}

\subsection{Experimental Setup}
Unless otherwise specified, the transceiver-related parameters are set as: $N_{t}=256$, $N_{r}=256$, $N_{h}=N_{v}=16$, $Q_{t_{\text{h}}}=Q_{v}=8$, $L_{h}=L_{v}=2$, $N_{D}=N_{t}N_{r}$. $t_{\text{max}}=100$ ps and $B_{t}=4$. The central frequency point is $f_{\text{c}}=100$ GHz, with $B=8$ GHz and $M=32$. The channel related parameters are: $\sigma_{\beta}=1$, $\sigma_{\alpha}=0.6$, $\tau_{\text{max}}=100$ ns, $K=3$.
The noise variances are $\sigma_{\text{c}}^{2}=\sigma_{\text{s}}^{2}=0$ dBm. The CSP-Net model employs the Adam optimizer, with a learning rate of 0.004 and a batch size $N_{b}$ of 16.
We define the SNR as $\frac{P_{t}}{\sigma_{\text{c}}^{2}}$ and the mean squared interval of angle (MSIA) between targets and the user as $\sqrt{\frac{\sum_{k=1}^{K}(\theta_{\text{s},k}-\theta_{\text{c}})^{2}+(\phi_{\text{s},k}-\phi_{\text{c}})^{2}}{2K}}$. 
The training data consist of communication and sensing channel data from various environments, generated based on the channel models in (\ref{equ:com_channel}) and (\ref{equ:sensing_channel}). The neural networks were trained on an NVIDIA RTX4060Ti, and the algorithms were tested in MATLAB 2024b. The system has 10 cores, a per-core clock speed of 2.5 GHz, and 160 floating-point operations per cycle.

\subsection{Comparison Methods}
The methods to be evaluated include:

\textbf{SA-Opt}: The squint-aware optimization-based benchmark for TTD-aided analog precoder design.

\textbf{CSP-Net}: The proposed unsupervised learning-aided scheme for analog precoder design without iteration.

\textbf{CBS-ISAC}: Controlling the beam squint to simultaneously cover the user and targets as described in Section 5.

\textbf{Opt w/o TTD}:  Alternating optimization of $\bm f_{\text{PS}}$ and $\bm p$ for the analog precoder design without TTDs.

\textbf{Comm-dedicated}: Focusing beams towards the user via delay-phase analog precoding for sub-THz systems in \cite{DPP}.

\textbf{Sensing-dedicated}: Spreading beams to cover targets via delay-phase analog precoding for sub-THz systems in \cite{squint_sensing}

\begin{table}[b]
    \vspace{-0.4cm}
    \centering
    \captionsetup{font=small}
    \caption{Ablation Experiments in CSP-Net}
    \vspace{-0.1cm}
    \begin{tabular}{p{1.7cm} |p{1.1cm} |p{1.9cm} |p{0.95cm} |p{1.15cm}}
    \hline
         &  Loss $\downarrow$ & Correlation $\uparrow$ & Rate $\uparrow$ & CRB $\downarrow$ \\
       \hline
       CSP-Net & \textbf{-1.915} & \textbf{0.9443}  & \textbf{18.52} &  \textbf{0.00154}    \\
       \hline
       real-valued & -1.802  &0.9167  & 18.34 & 0.00168     \\
       \hline
       w/o CBAM & -1.878 & 0.9388 & 18.37 & 0.00158     \\
       \hline
    \end{tabular}
    \label{tab:ablation}
    \vspace{-0.0cm}
\end{table}

\subsection{Comparative Study}

We first validate the training process of the proposed CSP-Net in Fig.~\ref{fig:training}. The overall loss decreases significantly during training and converges after around 10 epochs. Since the correlation is included in the customized loss function, the C-S channel correlation increases significantly, beneficial to improvement of C-S performance.
Through the comparison of the ablation experiments in Table.~\ref{tab:ablation}, the advantages of the complex-valued neural network and CBAM in processing channel data and integrating channel features can be verified respectively.

We then compare the ISAC performance of different schemes in Fig.~\ref{fig:rate_snr} and Fig.~\ref{fig:crb_snr}, under MSIA= 0.1. For communications, SA-Opt outperforms the CBS benchmark at high SNR with a gain of 2 dB, approaching the communication-dedicated scheme with perfect alignment for the user. In terms of sensing, the CRB of SA-Opt also outperforms CBS with a gain of 1.5 dB and approaches the sensing-dedicated scheme. This can be attributed to the increased C-S channel correlation, which enhances the multiplexing of resources for both functions. Furthermore, the significance of TTDs in tuning this correlation is evident when compared to the optimization without TTDs. It is worth noting that CSP-Net also outperforms existing benchmarks, approaching SA-Opt.

Furthermore, we compare these schemes under different user-target spatial distributions in Fig.~\ref{fig:rate_msia} and Fig.~\ref{fig:crb_msia}, fixing SNR as 10 dB. Due to the limited dispersion of beam squint, the array gain decreases as the user-target angular separation increases, and ISAC performance of benchmarks deteriorate drastically. However, the proposed schemes effectively counteract the performance loss by actively tuning the channels, suggesting a better adaptation to the spatial distribution. 

\begin{table}[b]
    \vspace{-0.4cm}
    \centering
    \captionsetup{font=small}
    \caption{Complexity comparison}
    \vspace{-0.1cm}
    \small
    \begin{tabular}{p{1.9cm}|p{6.2cm}}
    \hline
       Method  &  Complexity \\
       \hline
       SA-Opt  &  $\mathcal{O}(N_{\text{iter}}Q_{t}2^{B_{t}}(N_{t}^{3}N_{r}^{3}\!+\!2N_{t}^{2}N_{r}+2N_{t}^{2})\!+\!N_{\text{AO}}(M((2K)^{3}\!+\!12N_{t}^{2}K^{4})\!+\!M^{3}\log{1/\epsilon} )$       \\
       \hline
       CSP-Net  &  $\mathcal{O}(M^{2}(N_{t}^{2}+N_{t}N_{r})+Q_{t}^{2}+N_{t}^{2}+M^{2})$     \\
       \hline
       CBS-ISAC &  $\mathcal{O}(N_{\text{iter}}(N_{t}^{3}+Q_{t}^{3})\log{1/\epsilon})$      \\
       \hline
       Opt w/o TTD  &  $\mathcal{O}(N_{\text{AO}}(M((2K)^{3}\!+\!12N_{t}^{2}K^{4})\!+\!M^{3}\log{1/\epsilon} ))$      \\
       \hline
    \end{tabular}
    \label{tab:complexity}
    \vspace{-0.0cm}
\end{table}

\begin{figure}[t]
  \vspace{-0.0cm}
  \setlength{\abovecaptionskip}{-0cm} 
  \setlength{\belowcaptionskip}{0.1cm} 
  \centering
  \includegraphics[width=0.98\linewidth]{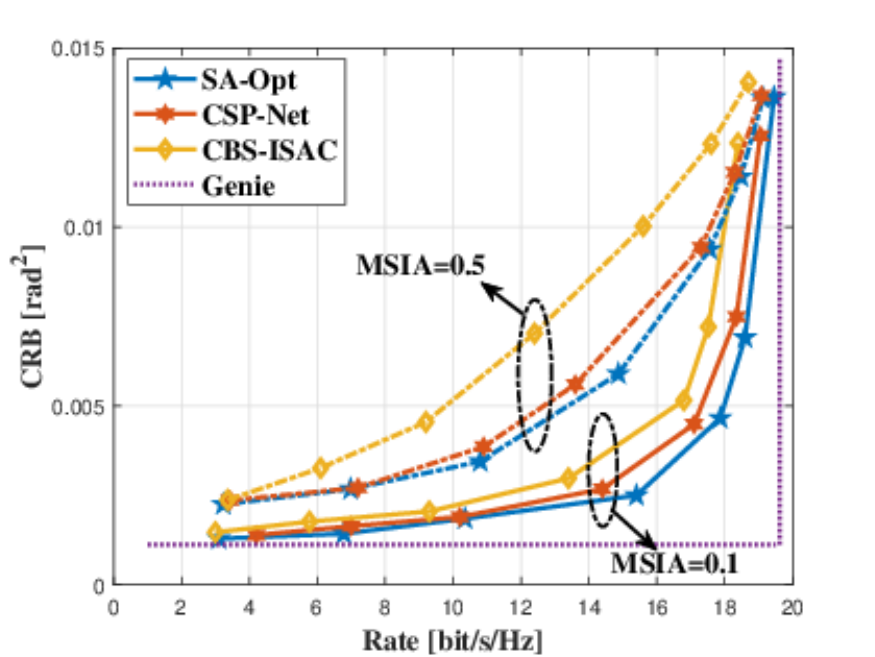}
  \captionsetup{font=small}
  \caption{ISAC performance boundaries. The SA-Opt benchmark achieves near-optimal ISAC performance, closely approaching the ideal case. The proposed CSP-Net scheme performs slightly worse than SA-Opt, with only a small gap. The squint-aware designs outperform existing methods.}
  \label{fig:bound}
  \vspace{-0.5cm}
\end{figure}

By comparing the ISAC performance boundaries achieved through these methods at SNR=10 dB in Fig~\ref{fig:bound}. Although communication-dedicated and sensing-dedicated schemes can each achieve optimal communication or sensing performance individually, they do not provide the dual-functional gain. The proposed schemes' effectiveness in enhancing dual-functional gain across different spatial distributions can be validated, since they expand the performance boundaries for dual functionality. Additionally, the performance boundaries are enhanced as the MSIA decreases, where a stronger correlation between the dual functionalities is established.

\begin{figure}[t]
  \vspace{-0.0cm}
  \setlength{\abovecaptionskip}{-0cm} 
  \setlength{\belowcaptionskip}{0.1cm} 
  \centering
  \includegraphics[width=0.98\linewidth]{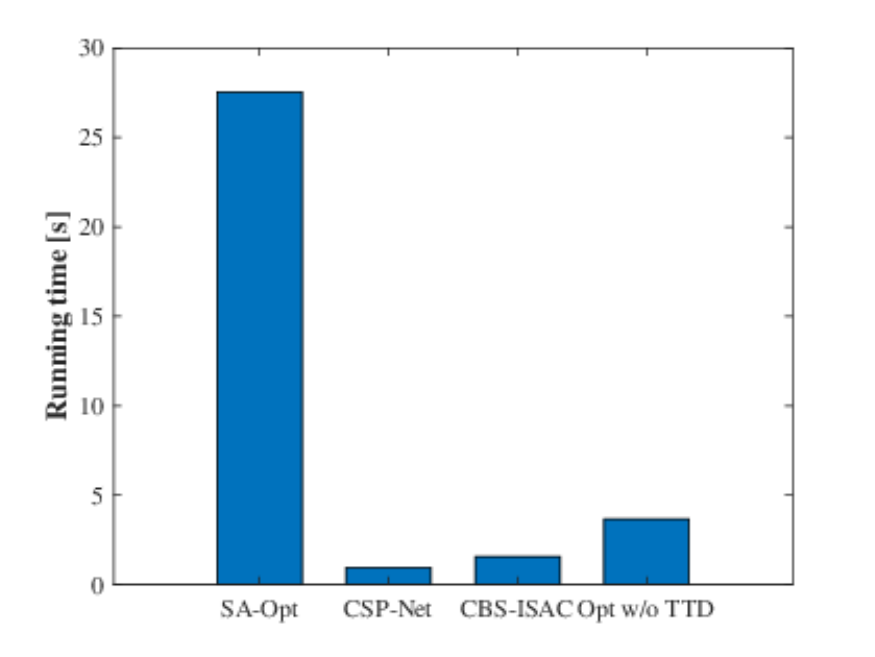}
  \captionsetup{font=small}
  \caption{Running time comparison. The SA-Opt benchmark requires the most design time, whereas the proposed CSP-Net exhibits the lowest computational complexity among these schemes.}
  \label{fig:time}
  \vspace{-0.5cm}
\end{figure}

The complexity of these approaches are analyzed in Table.~\ref{tab:complexity}, and the running time is compared in Fig~\ref{fig:time}. SA-Opt achieves the best ISAC performance among the methods at the cost of high complexity, while CSP-Net significantly reduces the complexity to be merely squared with $N_{t}$ and $M$, without a noticeable loss in performance. Therefore SA-Opt is well-suited for quasi-static environments, whereas CSP-Net is more effective in time-varying channels for mobile information networks.

\vspace{-0.2cm}
\section{Conclusion}
\vspace{-0.1cm}
In this paper, we proposed a novel analog precoder design to enhance ISAC performance in sub-THz systems. By leveraging the design degrees of freedom inherent in the RF front-end of sub-THz systems, we proactively adapt to the beam squint effect. Initially, we introduced an optimization-based method, SA-Opt, as a near-optimal benchmark. To further reduce the computational complexity, we presented a SoM-driven low-complexity alternative, CSP-Net. The proposed schemes effectively enhance the dual-functional gain and demonstrate robustness across various spatial distributions, highlighting the significant impact of C-S channel correlation on the ISAC performance boundaries. Furthermore, these schemes are compatible with the existing hardware architectures, offering promising applications for sub-THz systems.

\vspace{-0.2cm}
\appendix

\begin{figure*}[t]
     \begin{align}
    \xi_{\text{c},m,n_{1},n_{2}}\!&\approx \!\mathbb{I}(n_{1}=n_{2}\!=\!N_{r})\!+\!2\mathbb{I}(1\!\leqslant \!n_{1}\!\leqslant\! K,n_{2}\!=\!N_{r})\!\sum_{k=1}^{K}\!\mathbb{I}(\psi_{\text{s},k,m}\!=\!\psi_{\text{c},m})\!+\! \mathbb{I}(1\!\leqslant \!n_{1},\!n_{2}\!\leqslant\! K)\sum_{k_{1}}^{K}\!\sum_{k_{2}}^{K}\!\mathbb{I}(\psi_{\text{s},k_{1},m}\!=\!\psi_{\text{c},m})\mathbb{I}(\psi_{\text{s},k_{2},m}\!\!=\!\!\psi_{\text{c},m}),\nonumber  \\
    \vspace{-0.99cm}
    \xi_{\text{s},m,n_{1},n_{2}}\!&\approx \!\mathbb{I}(n_{1}\!=n_{2}\!=\!N_{r})\sum_{k=1}^{K}\mathbb{I}(\psi_{\text{c},m}\!=\!\psi_{\text{s},k,m})\vert\bm \Sigma_{m}[k,k]\vert^{2} \!+\!\mathbb{I}(1\leqslant n_{1},\!n_{2}\!\leqslant\!K)\sum_{k=1}^{K}\vert\bm \Sigma_{m}[k,k]\vert^{2}.
    \label{equ:xi}
    \tag{A.4}
    \vspace{-0.3cm}
\end{align}
\hrulefill
\vspace{-0.2cm}
\end{figure*}

\vspace{-0.1cm}
\section{Proof of the Proposition 1}
\label{app1}
According to \cite{IT_ISAC_Liufan}, the optimal $\bm R_{x,m}$ can be expressed as 
\vspace{-0.1cm}
\begin{equation}
\bm R_{x,m}^{\star}=\bm U_{m}\bm \Lambda_{m}\bm U_{m}^{\text{H}},
\vspace{-0.1cm}
\end{equation}
where $\bm U_{m}=[\bm A_{t,m}^{*}, \dot{\bm A}_{t,\theta,m}^{*},\dot{\bm A}_{t,\phi,m}^{*},\bm a_{t}(\theta_{\text{c}},\phi_{\text{c}},f_{m})]=[\bm u_{m,1}, \cdots, \!\bm u_{m,N_{r}}]\in \mathbb{C}^{N_{t}\times N_{r}}$ ($N_{r}\!=\!3K\!+\!1$) serving as the basis vectors under the given C-S channels $\bm h_{c}$ and $\bm G$, and $\bm \Lambda_{m}$ is positive semi-definite with $\bm \Lambda_{m}[n_{1},n_{2}]=\lambda_{m,n_{1},n_{2}}$. 
Then the achievable rate-CRB on the Pareto boundary is written as
\vspace{-0.25cm}
\begin{align}
    &R_{\text{c}}^{\star}=\sum_{m=1}^{M}\log_{2}\left[1+\frac{1}{\sigma_{c}^{2}}\sum_{n_{1}=1}^{N_{r}}\sum_{n_{2}=1}^{N_{r}}\lambda_{m,n_{1},n_{2}}\bm \Xi_{\text{c},m}[n_{1},n_{2}]\right],\nonumber \\
    \vspace{-0.4cm}
    &\text{CRB}^{\star}\approx N_{r}^{2}\left [\sum_{m=1}^{M}\sum_{n_{1}=1}^{N_{r}}\sum_{n_{2}=1}^{N_{r}}\lambda_{m,n_{1},n_{2}} \bm \Xi_{\text{s},m}[n_{1},n_{2}]\right ]^{-1},
    \vspace{-0.4cm}
    \label{equ:optimal}
\end{align}
$\bm \Xi_{\text{c},m}[n_{1},n_{2}]=\xi_{\text{c},m,n_{1},n_{2}}={\bm h_{\text{c},m}^{b}}^{\text{H}}\bm D_{t}^{\text{H}}\bm u_{m,n_{1}}\bm u_{m,n_{2}}^{\text{H}}\bm D_{t}\bm h_{\text{c},m}^{b}$, $\bm \Xi_{\text{s},m}[n_{1},n_{2}]=\xi_{\text{s},m,n_{1},n_{2}}=\bm u_{m,n_{1}}^{\text{H}}\bm A_{t,m}\bm \Sigma_{m}\bm \Sigma_{m}^{*}\bm A_{t,m}^{\text{H}}\bm u_{m,n_{2}}$ are both determined by the channels. By solving this multi-objective optimization, the optimal $\bm \Lambda_{m}$ can be written as
\vspace{-0.15cm}
\begin{equation}
    \bm \Lambda_{m}^{\star}=( \bm \Xi_{\text{c},m}+\gamma \bm \Xi_{\text{s},m})\frac{P_{t}}{M\text{Tr}(\bm \Xi_{\text{c},m}+\gamma \bm \Xi_{\text{s},m})}.
    \label{equ:Lambda}
    \vspace{-0.10cm}
\end{equation}
By substituting (\ref{equ:Lambda}) into (\ref{equ:optimal}), the rate and CRB at the Pareto optimum can be described via $\bm \Xi_{\text{c},m}$ and $\bm \Xi_{\text{s},m}$ as
\vspace{-0.1cm}
\begin{equation}
\begin{aligned}
    &\bm R_{\text{c}}^{\star}=\sum_{m}^{M}\log_{2}\left[1+\frac{1}{\sigma_{\text{c}}^{2}}\sum_{n_{1}}^{N_{r}}\sum_{n_{2}}^{N_{r}}\frac{P_{t}(\xi_{\text{c},m,n_{1},n_{2}}^{2}+\gamma \xi_{\text{s},m,n_{1},n_{2}}\xi_{\text{c},m,n_{1},n_{2}})}{M\text{Tr}(\bm \Xi_{\text{c},m}+\gamma \bm \Xi_{\text{s},m})}\right],\\
    &\text{CRB}^{\star}\approx N_{r}^{2}\left[\sum_{m}^{M}\sum_{n_{1}}^{N_{r}}\sum_{n_{2}}^{N_{r}}\frac{P_{t}(\gamma \xi_{\text{s},m,n_{1},n_{2}}^{2}+\xi_{\text{s},m,n_{1},n_{2}}\xi_{\text{c},m,n_{1},n_{2}})}{M\text{Tr}(\bm \Xi_{\text{c},m}+\gamma \bm \Xi_{\text{s},m})}\right]^{-1}. \nonumber
\end{aligned}
\vspace{-0.1cm}
\end{equation}
Leveraging the sparsity of beamspace channels \cite{shijian_MI_max}, we denote the indices of peaks in $\widehat{\bm h}_{\text{c}}^{b}$ and $\widehat{\bm h}_{\text{s}}^{b}$ as $\{\psi_{\text{c},m}\}$ and $\{\psi_{\text{s},k,m}\}$. The similarity of these peaks is defined as $\mathcal{S}_{m}(\bm h_{\text{c}},\bm G)=\sum_{k=1}^{K}\mathbb{I}( \psi_{\text{c},m}=\psi_{\text{s},k,m})$ and $\mathcal{S}(\bm h_{\text{c}},\bm G)=\sum_{m=1}^{M}\mathcal{S}_{m}$ with $\mathbb{I}(\cdot)$ being the indicator function. It can be proved that a higher $\text{Cor}(\bm h_{\text{c}},\bm G)$ gives rise to a higher $\mathcal{S}(\bm h_{\text{c}},\bm G)$ \cite{KL_proof}.
As derived in (\ref{equ:xi}) at the bottom of the previous page, as $\mathcal{S}(\bm h_{\text{c}},\bm G)$ increases, both $\xi_{\text{c},m,n_{1},n_{2}}$ and $\xi_{\text{s},m,n_{1},n_{2}}$ improve, leading to an increase in $R_{\text{c}}^{\star}$ and a reduction in $\text{CRB}^{\star}$ along the Pareto boundary. Therefore a higher C-S channel correlation improves the dual-functional gain.





\vspace{-0.25cm}
\bibliographystyle{IEEEtran}
\bibliography{IEEEabrv,myrefs}

\end{document}